\documentclass[aps,prb,superscriptaddress,showpacs,showkeys,twocolumn,amsmath,amssymb,preprintnumbers]{revtex4-1}

\usepackage{times}
\usepackage{color}
\usepackage{graphicx}
\usepackage{amsmath,amsbsy,amsfonts,mathrsfs}
\usepackage[version=3]{mhchem} 
\usepackage{dcolumn}
\usepackage{textcomp}

\begin{document}
\bibliographystyle{apsrev}

\preprint{doi:10.1016/j.susc.2015.10.052; Surf. Sci. 648, 256-261.}

\title{Role of Rotations on Surface Diffusion of Water Trimers on Pd\{111\}}

\date{\today}

\author{V\'ictor A. Ranea}

\affiliation{CCT-La Plata-CONICET, Instituto de Investigaciones
Fisicoqu\'imicas Te\'oricas y Aplicadas (INIFTA), Facultad de Ciencias
Exactas, Universidad Nacional de La Plata. Calle 64 y Diagonal 113, 1900
La Plata, Argentina}

\author{P.L. de Andres}

\affiliation{
Instituto de Ciencia de Materiales de Madrid - CSIC, Cantoblanco, ES-28049 Madrid and,
Donostia International Physics Center, P. Manuel de Lardizabal 4, 20018 Donostia, Spain}


\begin{abstract}
Diffusion barriers for a cluster of three water molecules on Pd(111) have been estimated from ab-initio Density Functional Theory. 
A model for the diffusion of the trimer based in rotations yields a simple explanation of why the cluster can diffuse faster than a single water molecule by a factor $\approx 10^{2}$.\cite{salmeron1}
This model is based on the differences between the adsorption geometry for the three monomers forming the cluster. One member interacts strongly with the surface and sits closer to the surface (d) while the other two interact weakly and stay at a larger separation from the surface (u). The trimer rotates nearly freely around the axis determined by the d monomer. Translations of the whole trimer imply breaking the strong interaction of the d monomer with the surface. Alternatively, thermal fluctuations exchange the actual monomer sitting closer to the surface with a lower energetic cost. 
Rotations around different axis introduce a diffusion mechanism where a strong interaction is kept along the diffusion path between the water molecule defining the axis of rotation and the Pd underneath. 
\end{abstract}

\keywords{
water ; monomer ; trimer ; water clusters ; diffusion ; rotation assisted ; Pd\{111\} ;  ab-initio ; density functional theory
}

\maketitle

\section{Introduction}
\paragraph{General}
Adsorption and diffusion of water molecules on metal surfaces play an important role in a series of phenomena such as catalysis, corrosion, energy production, storage, etc (e.g. see pioneering work by Somorjai and others\cite{somorjai1,somorjai2,somorjai3,somorjai4}). 
In order to understand these phenomena, it is essential to investigate the interactions established between water molecules and metallic surfaces.\cite{thiel,henderson,flores}

The water molecule has a permanent dipole that facilitates long-range dipole-dipole electrostatic interactions between molecules and with the image dipoles induced in the metal. 
In addition, hydrogen bonds between molecules play an important role in clustering processes. At the distances that are of concern here, these are made of electrostatic interaction between electronic densities, as shown by detailed quantum chemistry studies of the energy of interaction between water dimers.\cite{Ranea04}
Finally, oxygen atoms can establish chemical bonds with metal atoms in the surface, as seen by monitoring the redistribution of the electronic clouds in the separated systems.\cite{Michaelides.03} 


During the diffusion of water molecules on a metal surface, the formation of clusters of molecules by processes of growth and nucleation has been observed. 
In a series of elegant experiments Salmeron et al. have used scanning tunneling microscopy for the study of the diffusion of those aggregates of water molecules.\cite{salmeron1} Atomic resolution observations of diffusion of several clusters formed with one to six water molecules have been reported, and the number of water molecules in these clusters could be counted. The experimental resolution was not enough to elucidate their internal structure, but such information can be obtained by applying theoretical techniques based on ab-initio Density Functional Theory (DFT). Accuracy and credibility of these methods rest on their ability to provide a simple and reasonable physical explanation of experiments and in the agreement that can be achieved between theoretical predictions and experimental data. 

\paragraph{Experimental}
The experiments mentioned above yield an unexpected result. Clusters of two and three water molecules diffuse on the surface of Pd \{111\} faster than the water monomer by about $10^4$, and $10^2$ respectively.\cite{salmeron1} 
These factors have been measured at 40 K, and a typical frequency related to diffusion has been established from an Arrhenius plot as $10^{12}$ MHz. 
Therefore, we deduce from the experiment a reduction of at least a 10\% in the diffusion barrier for the water trimer with respect to the water monomer, in spite of the fact that the trimer displays extra interactions with the surface and has an adsorption energy that approximately doubles the one for the monomer.
We seek an explanation based on physical interactions computed by ab-initio techniques and different ways of facilitating the diffusion of these clusters. In this work we shall focus in the trimer.

\paragraph{Theoretical Model}
A single water molecule adsorbs near a top position at a height of $2.41$ {\AA}, and with an adsorption energy of $-0.26$ eV. Interactions between the H atoms and the metal surface make the adsorption position to be slightly off the symmetric atop by about $0.12$ {\AA} in agreement with previous results. 
This has been intepreted as mostly due to the interaction of the occupied 1$b_1$ molecular orbitals with the metal 
electronic states.\cite{Michaelides.03}
To perform a translation of the molecule this equilibrium configuration needs to be broken, hence giving rise to a diffusion barrier. 
We estimate this barrier as $0.13$ eV from the  difference with the adsorption energy over 
symmetric neighbouring sites located near the bridges. 
The experimental value obtained from an Arrhenius plot is $0.126 \pm 0.007$ eV.\cite{salmeron1} 
  
Regarding the internal structure of clusters of water molecules, theoretical calculations tell us that different components (monomers) adopt different structural positions in order to maximize external interactions between water molecules and the metalic surface, {\it and} the internal interactions among them (hidrogen bonds). For example, in the case of the water dimer a molecule adsorbs closer to the substrate while the other one sits in a higher position. For simplicity of notation, we shall call them down ({\em d}) and up ({\em u}). Using this information Ranea et al. have physically explained the faster diffusion of dimers over monomers.\cite{Ranea04} According to their model clusters made of two molecules are nearly free rotors around the axis defined by the monomer located closer to the surface. Diffusion takes place by the combined action of a thermal fluctuation bringing the two molecules in the dimer to a similar height {\it and} the concerted tunneling of the four protons to produce an exchange between the characters {\em d} and {\em u} of the two water molecules.
The total probability for such an event is computed as the product of the individual probabilities, i.e. the addition of individual contributions to the barrier, and the effective diffusion barrier for the dimer results lower than the one for the monomer in an amount compatible with the experimental observation.

\begin{figure}[htb]
\includegraphics[width=0.99\linewidth]{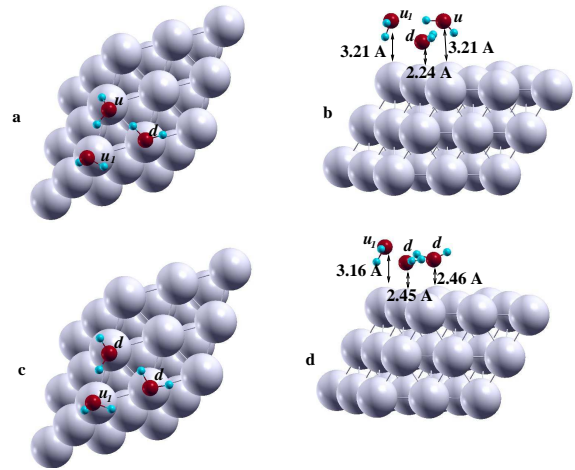}
\caption{\label{Fig1} \textbf{a} and \textbf{b} Top and lateral views of the
most stable water trimer adsorption configuration on the Pd\{111\} surface.
Adsorption energy = $-0.46$ eV.
The water monomer labeled {\em d} sits a short distance from the surface ($2.24$ {\AA}),
while the water
monomers labeled {\em u} ({\em u} and {\em u$_1$}) are far from the metal
surface ($3.21$ {\AA}). 
\textbf{c} and \textbf{d} Top and lateral views of the second best
equilibrium configuration. There are two water monomers labeled
{\em d} near the surface and one labeled {\em u} farther away. Adsorption energy = $-0.40$ eV.
Distances from the oxygen atoms to the nearest Pd atoms are shown in the
lateral views.
Big (silver), medium (red) and small (light blue) balls stand for Pd, O and H
atoms.} 
\end{figure}

In this paper, we extend these ideas to the water trimer. 
In the potential energy surface (PES) of the water trimer adsorbed on the
Pd\{111\} surface a minimum was found for a configuration with an 
energy of $-0.46$ eV. In this configuration one water molecule is
strongly bound to the Pd atom underneath it while the other two molecules forming the trimer stay over Pd atoms but at a
larger distance from the surface than the first molecule. By similarity with the dimer, we label the
three water molecules as {\em d}, {\em u} and {\em u'}, cf. 
Fig.~\ref{Fig1}. 
The O-Pd distances are $2.24$ \AA~ for {\em d}
and $3.21$ \AA~ for both {\em u} and {\em u'}. 
The molecular plane of the water
molecule labeled {\em d} is nearly parallel to the surface plane with the
hydrogen atoms only slightly higher from the
surface. The adsorption configuration of this monomer is similar to the
adsorption configuration of the isolated water molecule, but it is $0.17$ \AA~
closer to the surface. 
There is one H-bond between each of the pairs of molecules, while the other
hydrogen atoms are almost pointing to the surface in the molecules with the
labels {\em u}. 
The calculated adsorption energy for the
trimer is $-0.46$ eV, i.e. about $1.8$ times the adsorption energy of the single monomer. 
This comparison shows the importance of the H-bonds inside the
water trimer compared against the water-metal bonding. Comparison of the calculated
activation energies for surface diffusion of the adsorbates via traslation
(in any direction) shows that monomer diffusion is more likely than trimer
diffusion, since a stronger interaction needs to be broken. This result is in
disagreement with experimental results mentioned above. \cite{salmeron1}

As an alternative to translations we introduce a model where the water trimer
diffuses by rotation of the whole cluster around the {\em d} molecule. 
At some point during that rotation the trimer picks up a thermal fluctuation
to transform one of the two {\em u}
molecules into a {\em d} one (labeled {\em d'}), and later on the {\em d} molecule 
is transformed into a
{\em u}-like one (labeled {\em u'}).
This process has a kinetic barrier of 0.06 eV.
Rotations continue around the new {\em d'} axis and
the trimer is effectively translated without fully breaking
the water-metal interaction, therefore with a reduced diffusion barrier.
We shall come back to this mechanism in more detail later.
Notice that unlike the water dimer, the trimer is purely classical and the mechanism is based on thermal fluctuations only. 
Tunneling processes for the trimer involve too many protons, and their total mass reduces the likelihood of these processes below the values estimated for thermal processes.

\section{Methodology}

\paragraph{Density Functional Theory}
Total energies and diffusion barriers have been calculated using first-principles density functional theory (DFT).\cite{Kohn64,Kohn65}
The Vienna Ab initio Simulation Package (VASP) \cite{Kresse.94,Kresse.96} code
was used to investigate the water trimer diffusion on the Pd\{111\} surface.
The Kohn-Sham equations were solved using the projector augmented wave (PAW)
method \cite{Blochl.94,Kresse.99} and a plane-wave basis set including plane
waves up to 400 eV. Electron exchange and correlation energies were calculated
within the generalized gradient approximation (GGA) in the PBE form.\cite{pbe}
While total energies are converged to a precision better than $10^{-6}$ eV, the comparison of key values 
with an LDA functional, and the error bars in the experiments, allow us to estimate accuracy as $\pm 0.01$ eV.
Atoms are considered in equilibrium when forces are below $0.03$ eV/{\AA}.
El criterio de convergencia que utilic\'e es el normalmente usado, cada componente de
la fuerza aplicada a cada \'atomo que puede moverse es menor que 0.03 eV/\AA.
Values below those thresholds are considered indistinguishible in this paper.  
Van der Waals interactions increase adsorption energies by approximately 25\%.
Within the range of distances relevants for our problem  
we find that Van der Waals adds a nearly constant offset that cancels out in energy differences
and affects very little to the parameters determining diffusion rates.

\paragraph{Structure} 
The system (adsorbate + substrate + vacuum) is modelled by a rombohedral 
supercell with lattice constants: 
\textit{a} = $8.2519$, \textit{c} = $20.2130$\AA~, $\gamma=60^{\circ}$.\cite{Wyckoff}
The atomistic model for the Pd\{111\} surface consists of a slab formed by a
$3 \times 3$ two-dimensional cell parallel to the surface with a set of three
layers in the perpendicular direction. The vacuum separator is larger than
$10$ {\AA} defining the surface and preventing spurious interactions between
images in the periodic system. Water molecules have been adsorbed on only one
side of the slab. The two atomic planes located on the other side remain fixed
in the corresponding positions to a semi-infinite system, while atoms on the
last layer in contact with the adsorbates, and the adsorbates themselves, can freely relax
in all directions.
The first Brillouin zone was sampled with a ($3\times3\times1$) $\Gamma$
centered mesh.
A cubic cell with lattice constant $15$ \AA~ was used to calculate the optimized structures
and energies of the water monomer and trimer isolated species. 
Only the $\Gamma$ point was used for these clusters. 

The adsorption energy has been computed as:

\begin{equation} \label{eq:1}
E_{ad}=E(\textit{adsorbate}/\textit{srf})-E(\textit{adsorbate})-E(\textit{srf})
\end{equation}

The first term of equation (\ref{eq:1}) is the energy of the optimized
configuration of the adsorbate on the clean relaxed surface. The second term
of equation (\ref{eq:1}) is the gas phase energy of the isolated adsorbate.
The third term of equation (\ref{eq:1}) is the energy of the clean optimized
Pd\{111\} surface. With this definition stable configurations come as
negative values of $E_{ad}$.

\paragraph{Diffusion}
Similarly as in reference\cite{Ranea04} we compute diffusion rates as the product of a typical frequency, $w$,
giving the number of 
times the object is approaching the transition state times the Boltzmann factor giving the probability to pick up a thermal fluctuation with enough energy to overcome the barrier, $B$:
$$
D = w e^{-B/k_B T}
$$
\noindent
The value of $w$ can be estimated from typical phonon frequencies. In this expression the exponential function
dominates the behaviour as a function of the values of the barrier, $B$, or the temperature, $T$.
We have estimated $w$ from the frequencies for normal modes having amplitudes of vibration in
the direction of the diffusion path using a small cluster representative of the
interaction between the water molecule and the Pd atoms. These values have been computed with a
localized basis set of gaussians (cc-pVTZ\cite{ccpVTZ} for H and O, and sdd for Pd\cite{sdd}) 
and are in the order of $5$ to $10$ THz.\cite{gaussian}
We argue that this is an acceptable value according to the experimental value of $1$ THz 
since the experimental error of $\pm 7$ meV in the determination of the barrier
affects diffusion rates via the exponential term at $T=40$ K in factors between $0.1$ and $10$,
respectively. Therefore, the discrepancy in the prefactor 
is well inside the error bar for the energy of the transition state, can be absorbed on it,
and cancels out in relative
comparisons of different temperatures, or between different aggregates.

\section{Results.}


\paragraph{Adsorption geometry}
The cluster of three water molecules (trimer) interacting with the Pd\{111\} surface displays an equilibrium configuration with an adsorption energy of $-0.46$ eV. The trimer is centred on a hollow site, with the oxygen atoms located near Pd top positions. 
The difference in the calculated adsorption energies for the trimer
centered in the {\em fcc} and in the {\em hcp} hollow configurations is less than $0.01$ eV.

Interactions between the water molecules in the trimer dictate that one of them stay closer to the surface, at about $2.24$ {\AA} (water monomer labeled {\em d} in Fig. \ref{Fig1}\textbf{a} and \textbf{b}), while the other two sit at a larger distance of $3.21$ {\AA} (water monomers labeled {\em u} and {\em u'} in Fig. \ref{Fig1}\textbf{a} and \textbf{b}).
There is an attractive electrostatic interaction between each of the two water molecules labeled {\em u} or {\em u'}, and the surface. The absolute value of the adsorption energy, $0.46$ eV, is higher than the corresponding to the monomer, $0.26$ eV, by a factor $\approx 1.8$. Therefore, we interpret that the two u-like monomers account for about the same interaction energy as an isolated monomer. 
Two H-bonds keep the water trimer internally bound and are located near the plane defined by the three oxygen atoms. Two OH directions point away from that plane, towards the metallic surface. If these two are constrained to point away from the surface the interaction is weakened by $0.28$ eV.
We conclude that the two u-like members of the trimer do not form strong direct bonds between the oxygen atoms and the palladium, and interact with the surface mostly electrostatically.  
Therefore, the water trimer is bound to the Pd\{111\} surface via the lone pair of the oxygen atom of the water molecule labeled
{\em d} in Fig. \ref{Fig1}. 
It is interesting to notice that adsorption of a single water molecule (monomer) takes a geometrical configuration similar to the position of the member {\em d} in the trimer.
We are not taking into account Van der Waals like interactions since some parametrization would be needed that it would imply breaking the first-principles approach we have adopted. Van der Waals is expected not to alter our conclusions based on chemical and H-bonds interactions. Van der Waals interactions tend to favour maximum coordination configurations. Therefore, these would be more important near the hollows/bridges/tops, in that order.  

\begin{figure}[htb]
\includegraphics[width=0.99\linewidth]{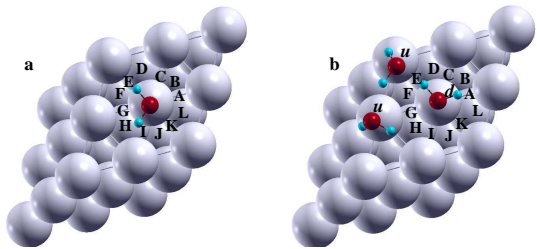}
\caption{\label{Fig2} \textbf{a} Top view of the most stable water monomer
adsorption on the Pd\{111\} surface. Adsorption energy = $-0.26$ eV.
The letters show the neighbour adsorption configurations also tested for the
water monomer adsorption. 
Adsorption energy = $-0.13$ eV for configurations A, E, G and K.
Adsorption energy = $-0.11$ eV for configurations B, C, D, F, H, I, J and L.
\textbf{b} Top view of a stable water trimer adsorption on the Pd\{111\}
surface (idem Fig. \ref{Fig1}\textbf{a}). In order to calculated the activation
energy for the trimer, the whole trimer was traslated to the neighbour sites.
The letters around the water monomer labeled {\em d} indicate where this monomer
was initially located. 
Adsorption energy = $-0.20$ eV for configurations B, E, F, J and K.
Big (silver), medium (red) and small (light blue) balls stand for Pd, O and H
atoms.} 
\end{figure}

Using the same conditions as for the trimer (i.e. the same set of pseudopotentials, 
exchange and correlation potential, and other parameters like the energy cutoff and the
k-mesh)
the adsorption of a water monomer on the Pd\{111\} surface is calculated on the
top site and on the twelve neighbour sites shown in Fig. \ref{Fig2}\textbf{a}.
The most stable adsorption configuration locates the oxygen atom of the water monomer near the top of a Pd atom, at a distance of $2.41$ \AA, with an adsorption energy of $-0.26$ eV.
The other adsorption configurations resulted in adsorption energies of $-0.13$ (A, E, G and K), and $-0.11$ (B, C, D, F, H, I, J and L) eV. 
In all the configurations, the plane of the water monomer is nearly parallel to the metal surface. 

\paragraph{Diffusion barriers}
The activation energy for diffusion of the water molecule from one top to the next one is obtained by calculating the energy difference between the high-symmetry sites in the pathway.\cite{ranea12} 
This process corresponds entirely to a translation of the object, and results in an activation energy for the monomer of $0.13$ eV.
The same mechanism for the trimer results in a barrier of $0.26$ eV; 
adsorption energies for sites B, E, F, J, and K were locally in equilibrium
with adsorption energy of $-0.20$ eV, while positions A, C, D, G, H, and I were unstable and moved spontaneously to the other sites. 
Therefore, a mechanism for diffusion based on translations predicts that the monomer should be faster than the trimer. This result is in disagreement with the experiments\cite{salmeron1} and calls for an alternative mechanism. 

\begin{figure}[htb]
\includegraphics[width=0.99\linewidth]{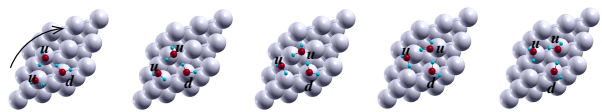}
\caption{\label{Fig3} Clockwise rotation of the water trimer around the axis defined by the monomer
{\em d}. Top view of the most stable water trimer adsorption configurations
on the Pd\{111\} surface for rotations of $0^{\circ}$ (Fig \ref{Fig1}\textbf{a}),
$15^{\circ}$, $30^{\circ}$, $45^{\circ}$ and $60^{\circ}$.
Adsorption energy = $-0.46$ eV. Every optimization was performed without
constrains but the atoms of the two deepest layers of the Pd\{111\} surface.
Big (silver), medium (red) and small (light blue) balls stand for Pd, O and H
atoms.} 
\end{figure}

We propose an alternative model for the surface diffusion of a water trimer based on a stationary configuration where the trimer adopts a configuration d d' u instead of the most stable d u' u. This configuration has an energetic cost of $0.06$ eV. It is interesting to notice that since both down-like monomers sit at the same height with respect to the surface it permits an exchange of the axis of rotation that results in a net translation of the trimer. Therefore, the mechanism relies on two key points. (i) The optimum configuration, d u' u, rotating quasi-barrierless around the axis through d, Fig.~\ref{Fig3}. (ii) A transition state with configuration d d' u, where d and d' sit at about the same distance from the Pd\{111\} surface. The lifetime of this metastable configuration is determined by the normal modes that tend to restore the optimum configuration. Frequencies for these normal modes have not been computed due to the complexity of the trimer, but we hypothesize that are in the same order as the ones obtained for the monomer. 

Figure \ref{Fig3} shows top views of several optimized configurations for the clockwise rotation of the water trimer around the {\em d} monomer.
The pictures are for rotations of $0^{\circ}$, $15^{\circ}$,
$30^{\circ}$, $45^{\circ}$ and $60^{\circ}$. 
This last configuration is
equivalent to the $0^{\circ}$ configuration.
Clockwise or anticlockwise rotations of {\em n} times $60^{\circ}$, where {\em n} is an integer number, around the {\em d} monomer, produce equivalent final configurations.
In all the calculations, all the atoms in the trimer and the atoms in the external layer of palladium are free to move accordingly to the calculated forces applied on them.
Only atoms in the two deepest layers of the Pd substrate have been constrained to their semi-infinite positions. 
All these configurations are stationary configurations in the PES.

\begin{figure}[htb]
\includegraphics[width=0.99\linewidth]{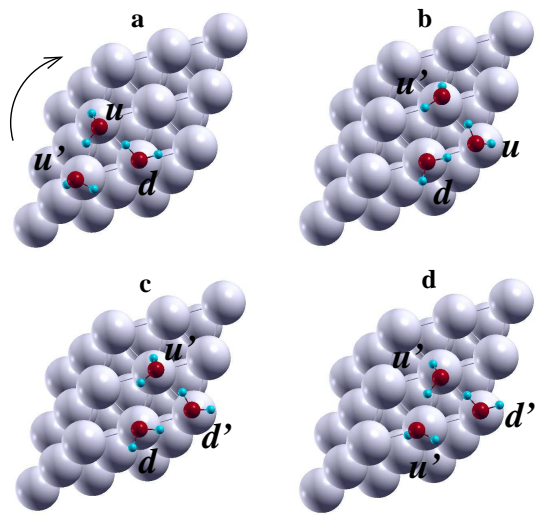}
\caption{\label{Fig4} Model of the water trimer translation via rotations with
an activation energy of $0.06$ eV.
\textbf{a} Water trimer adsorption in configuration {\em d}{\em u}{\em u'},
\textbf{b} Trimer clockwise rotation by $120^{\circ}$ around monomer {\em d},
\textbf{c} Monomer {\em u'} changes configuration to {\em d'} (endothermic
process by $0.06$ eV, water trimer in configuration {\em TS});
initial monomer {\em d} changs to {\em u'} (exothermic process by $0.06$ eV).
The trimer is again in an optimum configuration but the {\em d}-like member
has been translated by a unit cell vector.
Big (silver), medium (red) and small (light blue) balls stand for Pd, O and H
atoms.} 
\end{figure}

A simple example of how this mechanism produces the traslation of the trimer within this model is shown in Fig. \ref{Fig4} and is described here:

\begin{itemize}

\item
(a) We start with a trimer in an optimum configuration which water molecules
labeled {\em d},{\em u'} and {\em u} as shown in Fig.~\ref{Fig4}\textbf{a}.

\item
(b) We consider a clockwise rotation of $120^{\circ}$ around the monomer
{\em d} (i. e., two consecutive free rotations of $60^{\circ}$). This step is
quasi-barrierless. See Fig.\ref{Fig4}\textbf{b},

\item
(c) Monomer {\em u'} goes down to {\em d'}. This configuration
corresponds to the transition state with
{\em d},{\em d'},{\em u}. This process is endothermic by $0.06$ eV.
Thermal fluctuations (phonons) tend to transform any of the down-like
monomers into an upper-one with equal probability.
Let's assume the initial monomer {\em d} goes up to {\em u}.
The trimer has reached again an optimum configuration with
{\em u'},{\em d'},{\em u}. This process is exothermic by $0.06$ eV. 

\end{itemize}

\section{Conclusions}

Using ab-initio Density Functional Theory we have explained why the mere
translation of a water trimer cannot explain its experimental diffusion
rates on Pd\{111\}. 
We have introduced a model for diffusion where rotations of the trimer,
combined with picking up thermal fluctuations from surface phonons, rationalize
the low value determined for the effective diffusion barrier.

\section*{Acknowledgements.} 
This work has been financed by the MICINN, Spain (MAT2014-54231-C4-1-P).
We acknowledge the use of computing resources provided by the CTI-CSIC. 
This work was supported by Consejo de Investigaciones Cient\'ificas y T\'ecnicas
(CONICET) Universidad Nacional de La Plata, Argentina. 

\bibliography{trimH2Ov24ArXiv.bib}

\end{document}